\documentclass[flushrt]{emulateapj}

\usepackage{subfigure} 
\usepackage{epsfig}
\usepackage{xcolor}
\usepackage{tabularx}
\usepackage{booktabs}
\usepackage{multirow}
\usepackage{gensymb}

\newcommand{\npar}{\par \vspace{2.3ex plus 0.3ex minus 0.3ex}}

\usepackage{lscape}
\usepackage{longtable}

\begin{document}
\title{The evolution of metallicity and metallicity gradients from $z=2.7-0.6$ with KMOS$^{\mathrm{3D}}$} 


\author{Eva Wuyts\altaffilmark{1}, Emily Wisnioski\altaffilmark{1}, Matteo Fossati\altaffilmark{2,1}, Natascha M. F{\"o}rster Schreiber\altaffilmark{1}, Reinhard Genzel\altaffilmark{1,3,4},  Ric Davies\altaffilmark{1}, J. Trevor Mendel\altaffilmark{1}, Thorsten Naab\altaffilmark{5}, Bernhard R{\"o}ttgers\altaffilmark{5}, David J. Wilman\altaffilmark{2,1}, Stijn Wuyts\altaffilmark{6}, Kaushala Bandara\altaffilmark{1}, Alessandra Beifiori\altaffilmark{2,1}, Sirio Belli\altaffilmark{1}, Ralf Bender\altaffilmark{1,2}, Gabriel B. Brammer\altaffilmark{6}, Andreas Burkert\altaffilmark{5}, Jeffrey Chan\altaffilmark{2,1}, Audrey Galametz\altaffilmark{1}, Sandesh K. Kulkarni\altaffilmark{1}, Philipp Lang\altaffilmark{1}, Dieter Lutz\altaffilmark{1}, Ivelina G. Momcheva\altaffilmark{7}, Erica J. Nelson\altaffilmark{8}, David Rosario\altaffilmark{1}, Roberto P. Saglia\altaffilmark{1,2}, Stella Seitz\altaffilmark{2}, Linda J. Tacconi\altaffilmark{1}, Ken-ichi Tadaki\altaffilmark{1}, Hannah {\"U}bler\altaffilmark{1}, Pieter van Dokkum\altaffilmark{8}\footnotemark[*]}
\altaffiltext{1}{Max-Planck-Institut f\"{u}r extraterrestrische Physik, Giessenbachstr.~1, D-85741 Garching, Germany (evawuyts@mpe.mpg.de)}
\altaffiltext{2}{Universit{\"a}ts-Sternwarte M{\"u}nchen, Scheinerstr. 1, M{\"u}nchen, D-81679, Germany}
\altaffiltext{3}{Department of Physics, Le Conte Hall, University of California, 94720 Berkeley, USA}
\altaffiltext{4}{Department of Astronomy, Hearst Field Annex, University of California, Berkeley, 94720, USA}
\altaffiltext{5}{Max-Planck Institute for Astrophysics, Karl Schwarzschildstrasse 1, D-85748 Garching, Germany}
\altaffiltext{6}{Department of Physics, University of Bath, Claverton Down, Bath, BA2 7AY, UK}
\altaffiltext{7}{Space Telescope Science Institute, Baltimore, MD 21218, USA}
\altaffiltext{8}{Department of Astronomy, Yale University, P.O. Box 208101, New Haven, CT 06520-810, USA}

\footnotetext[*]{Based on observations obtained at the Very Large Telescope (VLT) of the European Southern Observatory (ESO), Paranal, Chile (ESO program IDs 092.A-0091, 093.A-0079, 094.A-0217, and 095.A-0047). This work is further based on observations taken by the 3D-HST Treasury Program (GO 12177 and 12328) with the NASA/ESA Hubble Space Telescope, which is operated by the Association of Universities for Research in Astronomy, Inc., under NASA contract NAS5-26555.}

\begin{abstract}
We present measurements of the [N~II]/H$\alpha$ ratio as a probe of gas-phase oxygen abundance for a sample of 419 star-forming galaxies at $z=0.6-2.7$ from the KMOS$^{\mathrm{3D}}$ near-IR multi-IFU survey. 
The mass-metallicity relation (MZR) is determined consistently with the same sample selection, metallicity tracer, and methodology over the wide redshift range probed by the survey. We find good agreement with long-slit surveys in the literature, except for the low-mass slope of the relation at $z\sim2.3$, where this sample is less biased than previous samples based on optical spectroscopic redshifts. In this regime we measure a steeper slope than some literature results. Excluding the AGN contribution from the MZR reduces sensitivity at the high mass end, but produces otherwise consistent results. There is no significant dependence of the [N~II]/H$\alpha$ ratio on SFR or environment at fixed redshift and stellar mass. 
The IFU data allow spatially resolved measurements of [N~II]/H$\alpha$, from which we can infer abundance gradients for 180~galaxies, thus tripling the current sample in the literature. 
The observed gradients are on average flat, with only 15 gradients statistically offset from zero at $>3\sigma$. We have modelled the effect of beam-smearing, assuming a smooth intrinsic radial gradient and known seeing, inclination and effective radius for each galaxy. Our seeing-limited observations can recover up to 70\% of the intrinsic gradient for the largest, face-on disks, but only 30\% for the smaller, more inclined galaxies. We do not find significant trends between observed or corrected gradients and any stellar population, dynamical or structural galaxy parameters, mostly in agreement with existing studies with much smaller sample sizes. In cosmological simulations, strong feedback is generally required to produce flat gradients at high redshift.   

\subjectheadings{galaxies: high-redshift, galaxies: abundances, galaxies: evolution} 
\end{abstract}

\section{Introduction}
\label{sec:intro}
The evolution of the cosmic star formation density and the mass growth of galaxies is largely driven by their available gas reservoirs, as determined from the interplay between gas accretion through cosmic inflow and mergers, star formation, and gas outflows driven by AGN and stellar feedback \citep[e.g.][]{Genel2008, Dekel2009, Bouche2010, Dave2012, Lilly2013, Tacconi2013}. The same processes also determine the metal content of galaxies, its spatial distribution and its evolution over cosmic time. Observationally, we find a tight correlation between stellar mass and gas-phase oxygen abundance in the local Universe, as robustly quantified in the Sloan Digital Sky Survey \citep[SDSS][]{Tremonti2004}. Theoretical models explain this relation through a combination of momentum- or energy-driven winds to remove metal-rich gas from the galaxy, inflows of metal-poor gas from the surrounding intergalactic medium, and variations in the star formation efficiency of galaxies \citep[e.g.][]{Dalcanton2004,Brooks2007,Finlator2008,Spitoni2010,Peeples2011,Lu2015}. Observational constraints on the mass-metallicity relation (MZR) and its evolution over cosmic time thus provide useful benchmarks for semi-analytic and numerical galaxy evolution models addressing the relative importance of gas inflow, enrichment, recycling and outflow. 

Towards higher redshift, an overall decrease in metallicity at fixed stellar mass has been well established with large samples since the recent developments of multi-object spectrographs and the availability of grism data from the Hubble Space Telescope \citep[e.g.][]{Erb2006,Zahid2011,Henry2013,Stott2013,Cullen2014,Steidel2014,me2014,Zahid2014,Sanders2015}. In absolute terms, the normalisation of the MZR evolution remains uncertain due to its dependence on the metallicity indicator used \citep{Kewley2008}. Since the set of observable rest-frame optical emission lines shifts with redshift, studies targeting different redshift ranges typically use different strong-line indicators, which complicates a direct comparison. Additionally, it remains unclear whether for the same indicator, the locally derived calibration remains valid at higher redshift given the different photo-ionisation conditions of high-z galaxies \citep[e.g.][]{Kewley2013,Steidel2014, Shapley2015}. New approaches using Bayesian analysis or $\chi^2$ minimisation for a larger set of emission lines to jointly solve for the galaxy metallicity as well as interstellar medium (ISM) properties such as ionisation parameter and electron density, are attempting to address this \citep{Perezmontero2014,Blanc2015}. 
\npar
In addition to the total metal content of galaxies, the spatial distribution of heavy elements within galaxies further constrains their baryonic and chemical assembly history. In the local Universe, abundance measurements of individual H~II regions generally find that galaxies have negative gradients, where the inner regions have higher metallicities than the outer disk regions, indicative of an inside-out growth scenario \citep{Zaritsky1994,vanZee1998,Sanchez2014}. There are indications of a flattening of the gradient outside the isophotal radius, possibly due to lower star formation efficiency in the outer disk \citep[e.g.][]{Goddard2011,Bresolin2012}. Mergers have been found to exhibit flat gradients, likely due to interaction-induced inflow of metal-poor gas in the centre and metal mixing \citep{Kewley2010,Rupke2010a,Rupke2010b,Perez2011,Sanchez2014}. At high redshift, dilution from gas infall into the galaxy centre has been invoked to explain some observations of positive abundance gradients \citep{Cresci2010}.

The time evolution of abundance gradients can be predicted from classical analytical models of the chemical evolution of the Milky Way disk \citep{Chiappini2001,Molla2005,Fu2009}, or from cosmological hydrodynamical simulations \citep{Kobayashi2011,Rahimi2011,Few2012,Pilkington2012,Gibson2013,AnglesAlcazar2014}. \cite{Pilkington2012} find little agreement in an extensive comparison of different simulations and chemical evolution models, which they trace back to the varying treatment of star formation and feedback in the simulations. The impact of the assumed feedback scenario is examined further by \cite{Gibson2013}. In this study, ``normal'' feedback (using 10-40\% of SN energy to heat the surrounding ISM) predicts steep gradients of -0.3~dex/kpc at $z\sim2$ which flatten over time, while ``enhanced'' feedback (including radiation pressure from massive stars) can better redistribute metal-enriched gas over large spatial scales and produces relatively flat gradients at all times. These studies highlight how robust gradient measurements at high redshift can provide powerful constraints on the uncertain nature of stellar feedback processes. 

 
Unfortunately also the observational measurements present challenges, requiring significant time investments to spatially resolve the fainter emission lines (typically [N~II] and/or H$\beta$) used in the strong-line metallicity calibrations. The current sample available in the literature consists of 90 gradients at $z=0.8-3.8$. Adaptive-optics (AO) assisted SINFONI observations have resulted in measured gradients for nine H$\alpha$-selected HiZELS galaxies at $z=0.8-2.24$ \citep{Swinbank2012} and 19 SINS/zCSINF galaxies at $z=1.4-2.4$ (F{\"o}rster Schreiber et al. 2016, in prep.). Without AO, there are 26 gradients at $z\sim1.2$ from MASSIV \citep{Queyrel2012} and 10 gradients at $z\sim3-4$ from AMAZE \citep{Troncoso2014}. Taking advantage of the multiplexing capabilities of KMOS at the VLT, \cite{Stott2014} present a first sample of 20 gradients at $z\sim0.8$ from HiZELS. There are concerns whether the coarse spatial resolution of seeing-limited observations, typically 5~kpc at $z\sim2$, can recover the intrinsic gradient \citep{Yuan2013}. AO-assisted or space-based observations of lensed galaxies can reach spatial resolutions down to 200-300~pc in the source plane. For 16 lensed galaxies with OSIRIS+AO data \citep{Jones2010, Yuan2011, Jones2013, Leet2015} and one arc with HST grism data from the Grism Lens-Amplified Survey from Space \citep[GLASS,][]{Jones2015}, some steeper gradients have been recovered, but most measurements are comparable to non-lensed samples at similar redshift.


\npar
In this work, we exploit the large redshift coverage of the KMOS$^{\mathrm{3D}}$ survey from $z=2.7$ to $z=0.6$ for a consistent study of the evolution of the MZR over five billion years around the period of peak star formation activity in the Universe. We use the spatially resolved information unique to our integral field spectroscopic survey to measure abundance gradients for 180 galaxies, thus tripling the total sample. The Paper is organised as follows. \S\ref{sec:sample} summarises the KMOS$^{\mathrm{3D}}$ sample selection, observations, and data reduction. The mass-metallicity relation and its dependence on the presence of an AGN, as well as on SFR and galaxy environment is analysed in \S\ref{sec:mzr}. In \S\ref{sec:grad} we present the radial abundance gradients, and address the effect of beam-smearing. \S\ref{sec:summ} provides a summary. Throughout this work, we adopt the \cite{Chabrier2003} initial mass function and a flat cosmology with $\Omega_M = 0.3$ and H$_0 = 70$\,km\,s$^{-1}$\,Mpc$^{-1}$.


\section{The KMOS$^{\mathrm{3D}}$ Sample}
\label{sec:sample}
\begin{figure*}
\centering
\includegraphics[width=\textwidth]{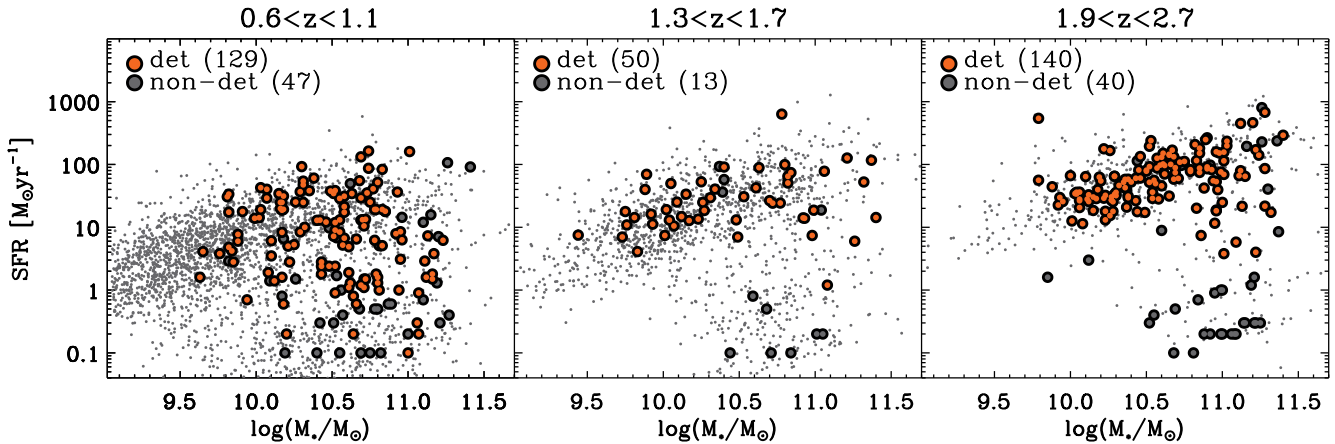}
\caption{Location of the observed KMOS$^{\mathrm{3D}}$ sample in the SED-derived SFR vs stellar mass plane in three redshift bins. The underlying grey dots represent the full mass-selected galaxy population at $z=0.6-2.7$ in the CANDELS fields. Our deep observations probe significantly below the main-sequence of star formation. \label{fig:sfrm}}
\end{figure*}

The galaxies analysed here are taken from the first two years of the KMOS$^{\mathrm{3D}}$ survey, which covers observations up to April 2015. The survey is described in detail by \cite{Wisnioski2015}. KMOS$^{\mathrm{3D}}$ is a five-year GTO survey with the multi-object near-IR integral field spectrograph KMOS at the VLT \citep{Sharples2013}, which aims to observe a mass-selected sample of $>600$ galaxies at $z=0.7-2.7$ in H$\alpha$ emission to study their spatially resolved kinematics and star formation. Targets are pre-selected to avoid OH skyline contamination based on a combination of existing spectroscopic campaigns and 3D-HST grism redshifts from the v4 catalogs \citep{Brammer2012, Momcheva2015} in the southern CANDELS fields UDS, COSMOS and GOODS-S \citep{Grogin2011,Koekemoer2011}. The availability of near-IR grism data avoids the bias towards bluer colours intrinsic in any selection based only on optical spectroscopic redshifts. Stellar masses, UV+IR star formation rates and global extinction values for the sample are derived from the CANDELS photometry \citep{Skelton2014} and available far-IR photometry \citep{Lutz2011, Magnelli2013}, using standard spectral energy distribution modelling with FAST \citep{Kriek2009} for a grid of \cite{bc03} models with exponentially declining star formation histories, \cite{Calzetti2000} dust extinction, and solar metallicity (see \citet{Wuyts2011} for details). Morphological parameters such as inclination and effective radius are available from GALFIT fitting of the F160W CANDELS data by \cite{vanderwel2012}. 
\npar
This work includes 419 targets, 176-63-180 each in the $YJ$, $H$ and $K$-bands respectively. The survey strategy deliberately focused on the $z\sim0.9$ and $z\sim2.3$ redshift slices first, only now starting to fill in the intermediate range at $z\sim1.5$ in $H$-band. Figure~\ref{fig:sfrm} situates the sample in the star formation vs stellar mass plane. Our deep integrations (baseline of 4-6-8hrs in $YJ$, $H$, and $K$ respectively) allow us to probe significantly below the main-sequence of star formation, though in this parameter space the detection fraction drops below our average of 76\%, especially in $K$-band. 

\npar
The data were reduced with the Software Package for Astronomical Reduction with KMOS \citep[\textsc{SPARK},][]{Davies2013}, complemented with custom IDL scripts for additional flat fielding, accurate centering and combining of data taken over multiple epochs, and PSF characterisation. Maps of the H$\alpha$ flux and kinematics are derived with the IDL emission line fitting code \textsc{LINEFIT}, which was originally developed for SINFONI data \citep{Forster2009, Davies2011}. We mask out all spatial pixels where the S/N in H$\alpha$ drops below 5, as well as outliers in velocity or velocity width or in their respective uncertainties. More details on the data reduction and kinematic mapping can be found in \cite{Wisnioski2015}. 

Guided by the KMOS$^{\mathrm{3D}}$ survey design which targets primarily H$\alpha$ emission, we derive gas-phase metallicities from the N2 index, the ratio of [N~II]~$\lambda$6583 to H$\alpha$ emission line fluxes. We note a number of known concerns related to this metallicity indicator: contribution of AGN and shock ionisation to the [NII] emission (further addressed in \S~\ref{subsec:agn}), saturation at high metallicities, variations in the N/O ratio and secondary nitrogen production, and contribution of the warm diffuse ISM. Given these possible biases, we show our results in terms of the observed [N~II]/H$\alpha$ ratio. When a conversion to metallicity is required for a comparison to the literature, the linear conversion by \cite{pp04} is used, $12+\mathrm{\log(O/H)} = 8.9+0.57 \times \mathrm{\log([N~II]/H\alpha)}$. This carries a systematic uncertainty of 0.18~dex.
 

\section{The Mass-Metallicity Relation}
\label{sec:mzr}
We create an integrated 1D spectrum for each galaxy by co-adding all the spatial pixels within the mask after velocity-shifting each spatial pixel to remove the imprint of the overall galaxy kinematics. The H$\alpha$ and [N~II] emission lines are jointly fit with a three-component Gaussian model, forcing a common redshift and line width, and constraining the [N~II]~$\lambda$6583/[N~II]~$\lambda$6548 doublet ratio to its theoretical value of 3.071 \citep{Storey2000}. Uncertainties are derived via a Monte Carlo approach where the spectrum is perturbed following its noise spectrum. From these fits we derive H$\alpha$-based SFRs, corrected for dust extinction using the SED-derived reddening and accounting for additional extinction of the nebular lines following \cite{Wuyts2013}. The integrated [N~II] emission is detected at $>3\sigma$ for 90\% of the sample, [N~II]/H$\alpha$ ratios are reported in Table~\ref{tab:properties} in the Appendix.

\subsection{AGN contamination}
\label{subsec:agn}

It is well known that the [N~II] emission can be contaminated by the ionising spectrum of an active galactic nucleus. There are a number of different AGN indicators in use in the literature, all of which suffer from biases and incompleteness. The ``classical'' indicators include X-ray emission, radio emission and mid-IR colours. Keeping in mind that the coverage and depth of the various surveys varies across the CANDELS fields \citep[see][for more details]{Genzel2014}, we find a typically low percentage of 7\% AGN in our KMOS$^{\mathrm{3D}}$ sample.
Surveys which also target [O~III] and H$\beta$ emission can use the optical BPT diagnostic diagram \citep{bpt} to identify the presence of AGN in their sample, and typically find another $\sim$10\% contamination \citep{Zahid2014, Sanders2015}.  
The KMOS$^{\mathrm{3D}}$ survey currently does not cover the [O~III] or H$\beta$ line for the majority of its targets, but [N~II]/H$\alpha$ alone also serves as a useful diagnostic to identify AGN from rest-frame optical emission lines at high redshift \citep{Stasinska2006, Cid2011, Coil2015}. We find that 8\% of targets in our sample lie above a threshold [N~II]/H$\alpha > 0.6$, using the threshold from \cite{Kewley2001}.

The spatially resolved IFU data give access to two additional AGN indicators. The AGN contribution to the line emission is expected to be largest in the nuclear region of the galaxy, and possibly diluted by the emission from the outer disk. We therefore create a nuclear spectrum for each of our targets by co-adding only the central 0.4\arcsec. The nuclear [N~II]/H$\alpha$ ratio lies above 0.6 for an additional 5\% of the sample. Secondly, \cite{Forster2014} and \cite{Genzel2014} reported evidence for nuclear AGN-driven outflows of ionised gas based on the presence of a broad underlying component with FWHM=500-2000~km/s in the line profile of H$\alpha$ and the forbidden lines. We visually examine all nuclear spectra as described by \cite{Genzel2014} and find evidence of underlying broad emission for 27\% of the sample analysed in this paper. The incidence of each of the AGN indicators described above is a strong function of stellar mass, such that at $\log(M_*/\mathrm{M}_\odot)>10.9$, 70\% of the sample satisfies one or more of the AGN indicators. 

\begin{figure*}
\centering
\includegraphics[width=\textwidth]{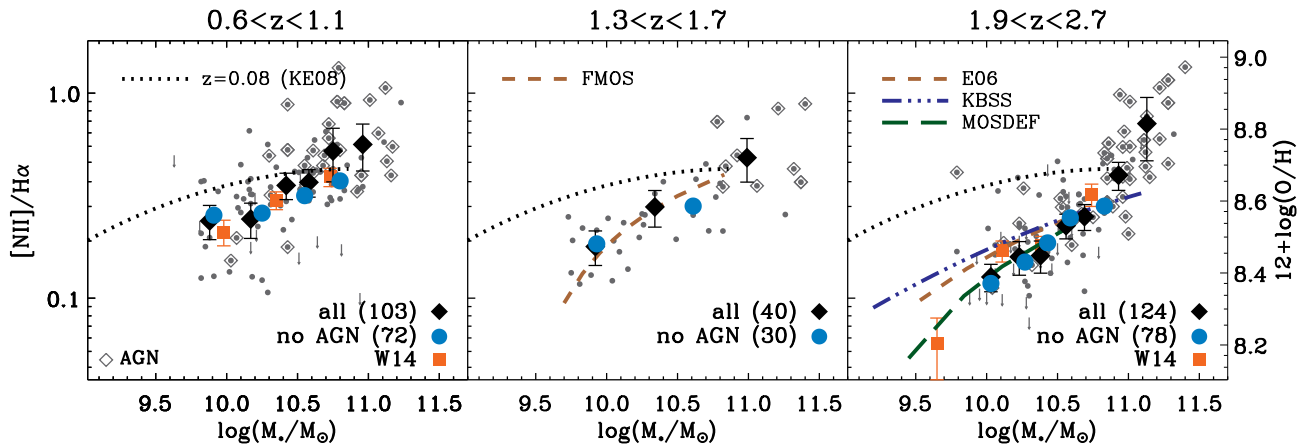}
\caption{Galaxy-integrated [N~II]/H$\alpha$ ratios as a function of stellar mass for the KMOS$^{\mathrm{3D}}$ sample. The right axis expresses galaxy abundance as $12+\log(O/H)$ as derived from the linear relation by \cite{pp04}. The grey datapoints in the background show our individual detections and upper limits, AGN are surrounded by an open diamond. The large black diamonds and blue circles correspond to the stacked spectra with and without AGN-identified targets respectively. The local relation from \cite{Kewley2008} is shown with a dotted line in each panel. The orange squares correspond to the earlier KMOS+SINFONI+LUCI results from \cite{me2014}. Other studies in the literature include the FMOS survey at $z\sim1.6$ \citep[brown dashed]{Zahid2014}, and at $z\sim2.3$ the sample from \cite{Erb2006} (brown dashed), the KBSS survey \citep[dark blue dot dashed]{Steidel2014} and the MOSDEF survey \citep[dark green dashed]{Sanders2015}. Their MZR relations are represented consistently based on metallicities derived from the N2 indicator and the linear \cite{pp04} conversion. \label{fig:mzr}}
\end{figure*}

\subsection{Stacking}
The KMOS$^{\mathrm{3D}}$ sample has a high detection rate of 90\% for the galaxy-integrated [N~II] emission, not unsurprising given the required depth to spatially resolve the H$\alpha$ emission. Nevertheless, we stack our galaxies in bins of stellar mass to robustly take into account the upper limits. Similarly to \cite{me2014}, we use inverse variance weighting of continuum-subtracted, velocity-corrected and H$\alpha$-normalized integrated spectra. The line fluxes and uncertainties are derived as the jackknife mean and standard error and reported in Table~\ref{tab:mzr}. We aim for 15-20 targets per bin. Figure~\ref{fig:mzr} shows the mass-metallicity relation for the stacked spectra with and without the AGN identified in the previous section. The strong dependence of AGN incidence on stellar mass results in a loss of many of the highest mass galaxies. However, in the overlapping mass regime at $\log(M_*/\mathrm{M}_\odot)\lesssim10.8$, both stacked relations are consistent within the uncertainties, suggesting that in this mass range the effect of the AGN contribution to the galaxy-integrated [N~II] emission is limited. At higher masses, the integrated [N~II]/H$\alpha$ ratios are often biased high.
\npar
The results are in excellent agreement with the MZR presented in \cite{me2014} based on the first year KMOS$^{\mathrm{3D}}$ data, augmented with observations with SINFONI at the VLT and the long-slit near-IR spectrograph LUCI at the Large Binocular Telescope on Mount Graham, Arizona. The latter two datasets are responsible for the extension to lower stellar masses, especially in $K$-band. Further comparison to the literature at $z\sim$1.6 is available from the FMOS survey, for 168 galaxies in the COSMOS field observed with the multi-object near-IR spectrograph FMOS on Subaru \citep{Zahid2014,Silverman2015}. The wide field of view surveyed includes many more very massive targets compared to KMOS$^{\mathrm{3D}}$, but the lack of spatially-resolved data limits the detection of AGN contributions, which could explain the somewhat higher abundances at the high mass end, where we have shown AGN to bias the [N~II]/H$\alpha$ ratios high. Most high-z emission-line science has so far been carried out at $z\sim2.3$, where the full suite of rest-frame optical diagnostic lines can be observed in the near-IR wavelength range. The MZR presented by \cite{Erb2006} was long the only study with a large sample ($\sim100$) at this redshift, albeit with few individual [N~II] detections. Recently, the new multi-object near-IR spectrograph MOSFIRE on Keck has changed the landscape with two large emission-line surveys of many hundreds galaxies each. The MOSFIRE Deep Evolution Field (MOSDEF) survey follows a similar target selection to KMOS$^{\mathrm{3D}}$ in the 3D-HST CANDELS fields observable from Mauna Kea \citep{Kriek2015}. The Keck Baryonic Structure Survey (KBSS) targets 15 quasar fields with a large sample of spectroscopically confirmed redshifts at $1.5 \lesssim z \lesssim 3.5$ and a wealth of multi-wavelength ancillary data \citep[e.g.][]{Steidel2004, Rudie2012, Steidel2014}. We have taken care to consistently compare relations which have been derived with the same N2 indicator and linear \cite{pp04} conversion employed here. Our results are in perfect agreement with the MOSDEF relation \citep{Sanders2015}, which is reassuring given our almost identical parent sample. The MZR relation reported by KBSS has a shallower slope, which is likely due to differences in the selected target population (Strom et al. 2016, in prep.).

\begin{table}
\centering
\caption{Stacked mass-metallicity relation in three redshift intervals, with and without AGN. \label{tab:mzr}}
\medskip
\begin{tabularx}{9cm}{lcc|cc}
                  & \multicolumn{2}{c}{\textbf{all}} &  \multicolumn{2}{c}{\textbf{no AGN}} \\ \toprule
Redshift & $\log(M_*/\mathrm{M}_\odot)$ &  [N~II]/H$\alpha$ & $\log(M_*/\mathrm{M}_\odot)$ & [N~II]/H$\alpha$ \\ \midrule
\textbf{$z\sim0.9$}   &   9.88   & $0.186\pm0.026$  &   9.91 & $0.218\pm0.043$ \\
	            & 10.17    & $0.250\pm0.065$ & 10.25 & $0.277\pm0.044$ \\
                    & 10.42    & $0.338\pm0.068$ & 10.55 & $0.364\pm0.072$ \\
                    & 10.58    & $0.436\pm0.049$ & 10.80 & $0.384\pm0.038$ \\
                    & 10.75    & $0.404\pm0.067$ &           &                              \\
                    & 10.96    & $0.432\pm0.122$ &           &                              \\ \midrule
\textbf{$z\sim1.5$}   &   9.92   & $0.165\pm0.026$  &   9.93 & $0.164\pm0.034$ \\
	            & 10.34    & $0.233\pm0.040$ & 10.61 & $0.262\pm0.057$ \\
                    & 10.99    & $0.368\pm0.204$ &           &                              \\ \midrule
\textbf{$z\sim2.3$}   & 10.03   & $0.119\pm0.016$  & 10.03 & $0.114\pm0.012$ \\
	            & 10.23    & $0.122\pm0.026$ & 10.27 & $0.134\pm0.024$ \\
                    & 10.38    & $0.142\pm0.017$ & 10.43 & $0.165\pm0.020$ \\
                    & 10.56    & $0.211\pm0.032$ & 10.59 & $0.231\pm0.034$ \\
                    & 10.69    & $0.293\pm0.062$ & 10.83 & $0.328\pm0.115$ \\
                    & 10.93    & $0.351\pm0.049$ &           &                              \\
                    & 11.13    & $0.542\pm0.400$ &           &                              \\ \bottomrule
\end{tabularx}
\end{table}


\subsection{The Effect of Star Formation Rate}
In the local Universe, \cite{Ellison2008} first reported an anti-correlation between metallicity and specific star formation rate (sSFR). Further studies showed how including the SFR as a second parameter in the mass-metallicity relation reduces the scatter \citep{Mannucci2010, Andrews2013}. There have been claims that this ``fundamental'' relation (FMR) between stellar mass, SFR and galaxy abundance is independent of redshift, such that the lower abundances observed at high redshift at fixed stellar mass are fully explained by the overall higher SFR at these epochs \citep[e.g.][]{Mannucci2010, Belli2013, Henry2013, Stott2013}. However, the majority of high-z studies report no dependence of the MZR on star formation rate, or an offset from the local FMR \citep{Zahid2014,me2014,Steidel2014,Troncoso2014,Sanders2015,Grasshorn2015}

\begin{figure}
\centering
\includegraphics[width=8cm]{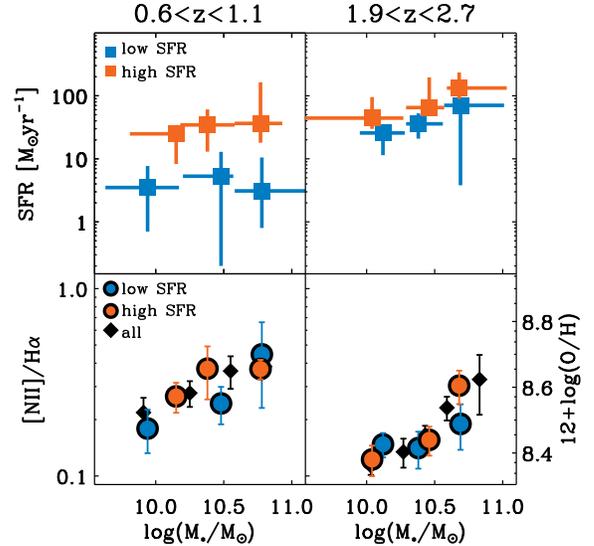}
\caption{The effect of star formation on metallicity, using the dust-corrected SFRs derived from the integrated H$\alpha$ flux. Top panels show the median value and range of a low and high SFR bin for three stellar mass bins at $z\sim0.9$ and $z\sim2.3$. The MZR for the low and high SFR bins separately is shown in the bottom panels, and compared to the MZR for all galaxies from Figure~\ref{fig:mzr}. There is no significant offset between the SFR bins. \label{fig:fmr}}
\end{figure}

Our data does not show a significant trend between inferred metallicity and sSFR, in contrast to recent results from the KMOS Redshift One Spectroscopic Survey \citep[KROSS][]{Magdis2016}. This could be due to a less complete removal of AGN contamination from their sample as well as a difference in survey depth at $z\sim1$ (KMOS$^{\mathrm{3D}}$: $\sim5$hrs; KROSS: $\sim2.5$hrs). To investigate the dependence on SFR at fixed mass and redshift, we split our AGN-excluded sample at $z\sim0.9$ and $z\sim2.3$ in two bins of SFR, using the dust-corrected SFRs derived from the integrated H$\alpha$ flux (Figure~\ref{fig:fmr}). The intermediate redshift bin at $z\sim1.5$ does not yet contain sufficient targets. We do not see a significant offset between the MZR separately derived for the low and high SFR bin. This does not change when we use the photometrically based SFR estimates from UV+IR broad-band data. The dynamic range of the $z\sim2.3$ sample is very limited, but over the SFR range probed at $z\sim0.9$ in KMOS$^{\mathrm{3D}}$, a significant effect on metallicity is seen in the local Universe \citep{Andrews2013}.


The equilibrium or ``gas-regulator'' picture of galaxy evolution driven by the balance between gas accretion, star formation and gas outflow, naturally finds a fundamental relation between stellar mass, SFR and metallicity \citep[e.g.][]{Lilly2013}. The lack of observed correlation remains puzzling, and could signify departures from the first-order assumptions of these simple analytical equilibrium models. First, however, the possible caveats related to metallicity measurements based on strong-line indicators should be more completely understood.




\subsection{The Effect of Environment}
Local density can have a strong effect on the overall evolution of galaxies. Specifically for the galaxy abundance, a denser environment could remove metal-poor gas from the outskirts of galaxies or result in shorter gas recycling times \citep{Oppenheimer2008}, enhancing the observed metallicity. However, offsets in the local MZR of field and cluster star-forming galaxies are observationally and theoretically constrained to $<0.05$~dex \citep{Ellison2009, Dave2011, Scudder2012, Hughes2013}. The few studies carried out to date in proto-cluster environments at $z\sim2$ mostly limit any enhancement of the abundance to $\le 0.15$~dex at $<2\sigma$ significance \citep{Kulas2013, Shimakawa2015, Kacprzak2015}; \cite{Valentino2015} in contrast find a metal deficiency of 0.25~dex at 4$\sigma$ significance for a cluster at $z=1.99$.

\begin{figure*}
\centering
\includegraphics[width=\textwidth]{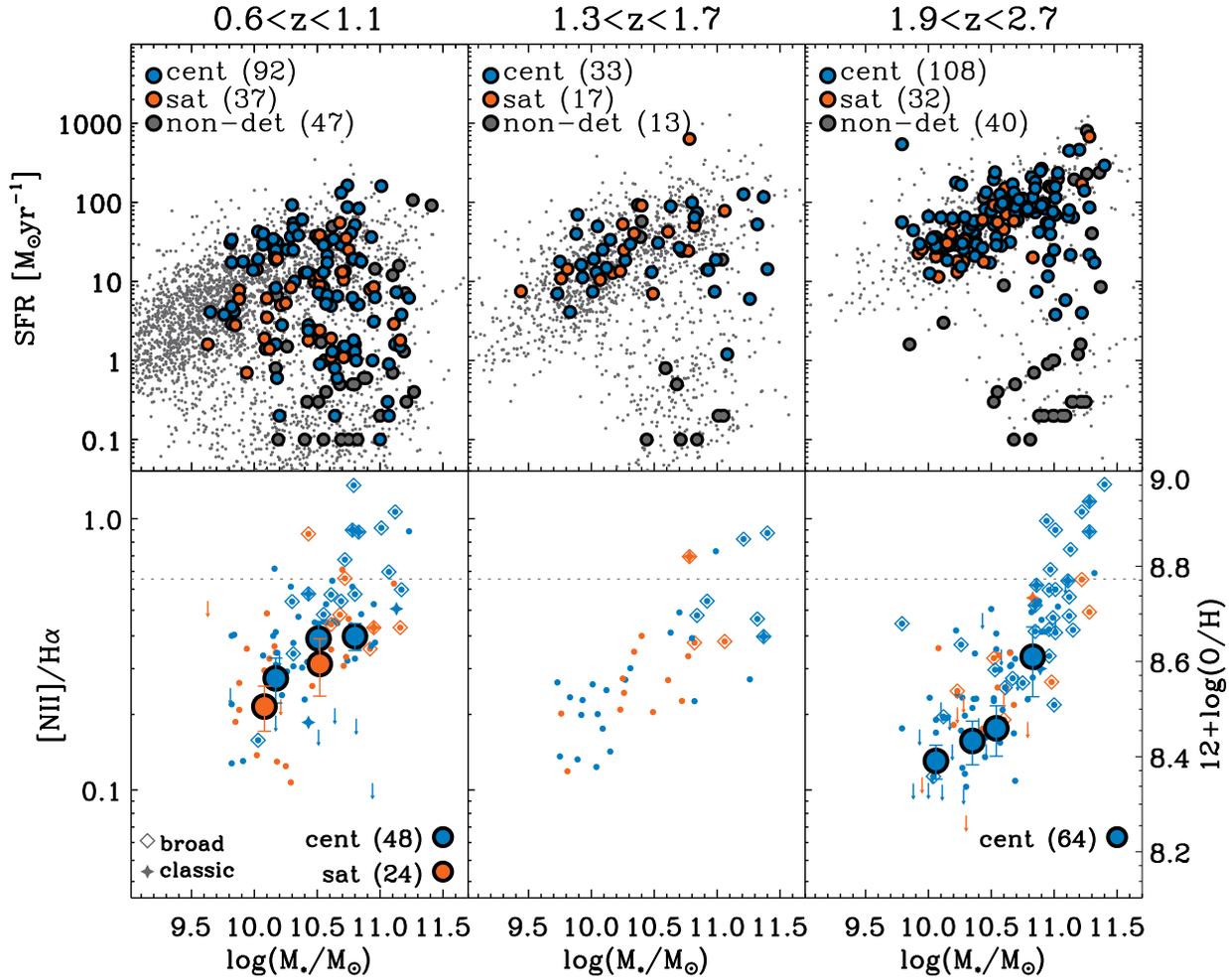}
\caption{\textit{(Top)} SFR vs stellar mass for the KMOS$^{\mathrm{3D}}$ sample (as in Figure~\ref{fig:sfrm}), color-coded as central and satellite galaxies in blue and orange respectively. We find the expected $\sim30$\% of satellites \citep{Fossati2015} at $z\sim0.9$ and $z\sim1.5$, at $z\sim2.3$ the focus on massive galaxies results in fewer satellite galaxies. Both subsets occupy a similar phase space in this diagram. \textit{(Bottom)} Galaxy-integrated [N~II]/H$\alpha$ ratios vs stellar mass (as in Figure~\ref{fig:mzr}), color-coded as central and satellite galaxies in blue and orange respectively. Only at $z\sim0.9$ do we have enough satellites for a stacked mass-metallicity relation excluding AGN. Central galaxies exhibit marginally higher [N~II]/H$\alpha$ ratios at 0.06~dex. \label{fig:env}}
\end{figure*}

With the KMOS$^{\mathrm{3D}}$ sample we can for the first time probe the effect of variations in local density on galaxy abundances for a large sample at high redshift. We characterise each KMOS$^{\mathrm{3D}}$ target as a central or satellite galaxy based on a determination of its mass-rank in an adaptive aperture and a comparison to simulations \citep[][Fossati et al. 2016 in prep.]{Fossati2015}. The expected $\sim30$\% satellite galaxies is roughly recovered, except at $z\sim2.3$ due to the stronger focus on massive galaxies in this redshift slice in the first two years of KMOS$^{\mathrm{3D}}$ target selection. Central and satellite galaxies occupy a similar phase space in the SFR - stellar mass plane, as shown in the top panel of Figure~\ref{fig:env}. At $z\sim0.9$, the sample contains enough non-AGN, [N~II] detected satellite galaxies to stack both categories separately in bins of stellar mass and compare the MZR. We find a metallicity enhancement of 0.06~dex for central galaxies, though at low significance ($<2\sigma$) given the uncertainties. 

\section{Abundance Gradients}
\label{sec:grad}
The spatial distribution of the [N~II]/H$\alpha$ ratio across individual galaxies contains additional information beyond a simple measurement of the galaxy-integrated abundance. For each target, a set of elliptical annuli is created, centred on the continuum light, angled along the kinematic position angle \citep[see][]{Wisnioski2015} and with an ellipticity matching the outer H$\alpha$ contours. Given the FWHM of our seeing-limited data, we ideally choose apertures with a width of 0.4\arcsec, or 2 spatial pixels. However, only 40\% of the sample has enough signal-to-noise over a large enough area to detect [N~II] at $>3\sigma$ in three subsequent 2 pixel-wide apertures, which is our requirement to robustly measure a radial gradient. Especially in the highest redshift bin at $z\sim2.3$, most galaxies are too small and for only 17\% can a gradient be measured. Therefore, our default apertures have a width of 0.2\arcsec. Where possible, we have checked that gradients measured with 0.2\arcsec\ and 0.4\arcsec\ wide apertures are consistent within the uncertainties. In each aperture, we create a velocity-corrected integrated spectrum and jointly fit the [N~II] and H$\alpha$ emission as described in Section~\ref{sec:mzr}. A linear fit of the [N~II]/H$\alpha$ ratio as a function of the semi-major axis radius (defined in the middle of each annulus) results in a measurement of the gradient $\Delta$N2/$\Delta$r in dex/kpc. Figure~\ref{fig:n2grad_examples} in the Appendix shows the H$\alpha$, velocity and [N~II]/H$\alpha$ map as well as the [N~II]/H$\alpha$ ratio as a function of radius for five example galaxies to illustrate the process. We have visually examined each gradient measurement to avoid catastrophic failures in the fit.

\begin{figure}
\centering
\includegraphics[width=8cm]{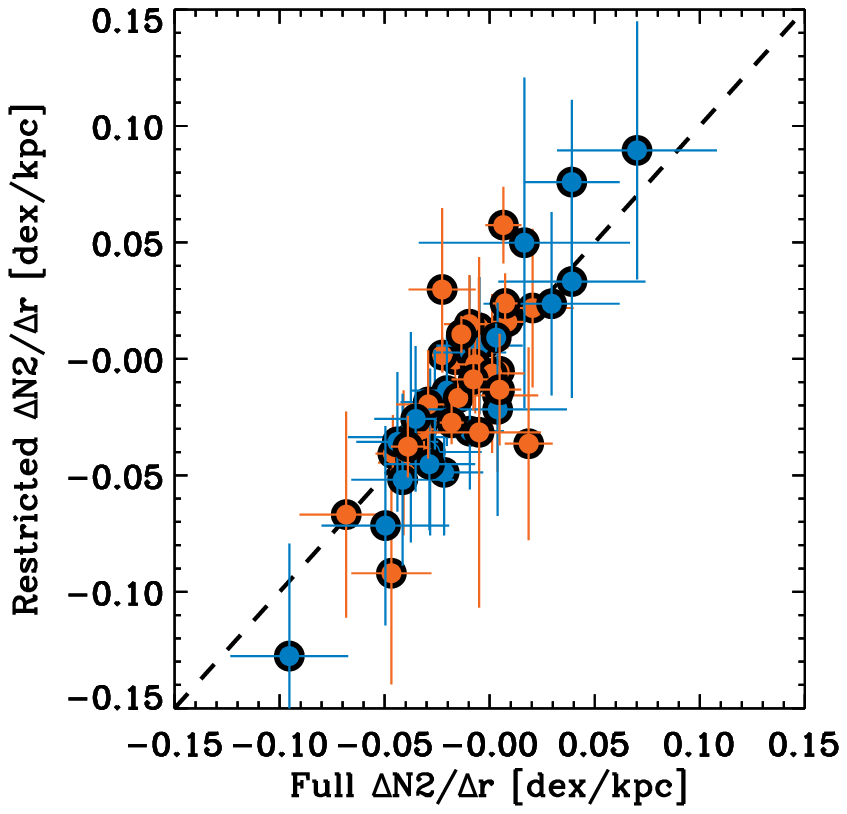}
\caption{A comparison of the full radial gradient with the gradient restricted to the outer disk for the galaxies flagged as AGN. For the orange symbols, the inner 0.4\arcsec\ covered by the two innermost apertures is excluded from the fit, for the blue symbols we can exclude only the innermost 0.2\arcsec\ to keep at least three radial [N~II] detections. The uncertainties for the restricted gradient on the y-axis are naturally larger due to the fewer radial data points, but on average the AGN does not significantly steepen the derived linear gradient.\label{fig:grad_agn}}
\end{figure}

For the galaxies identified as having some AGN contribution from one of the indicators described in Section~\ref{subsec:agn}, the [N~II] emission in the central region could be enhanced by the AGN and thus not accurately represent the metallicity of the ionised gas. In these cases, we exclude the inner two apertures (i.e. the inner 0.4\arcsec) from the radial fit. When only four apertures with robust [N~II] detections exist, only the innermost aperture is excluded. For these targets, removing two apertures and fitting a gradient to only the remaining outer two radial bins gives consistent results. AGN with only three radial [N~II] detections are excluded. For the full sample of AGN, the restricted gradients are on average $0.001\pm0.02$~dex/kpc flatter, i.e. fully consistent with the full radial gradients, as can be seen in Figure~\ref{fig:grad_agn}. 

\begin{figure}
\centering
\includegraphics[width=8cm]{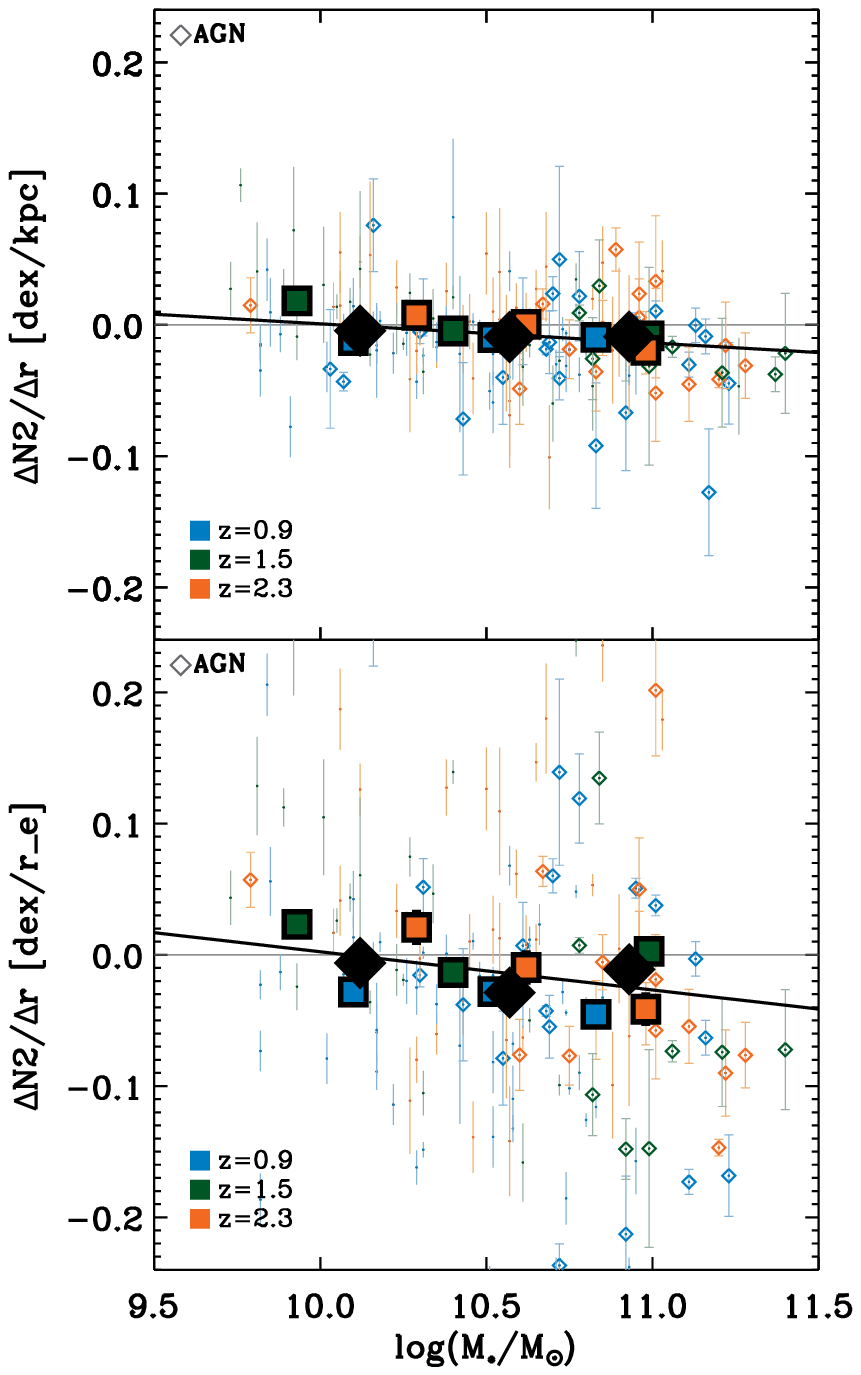}
\caption{Observed metallicity gradients as a function of stellar mass, color-coded by redshift. The \textit{top} panel shows gradients in [dex/kpc], in the \textit{bottom} panel they have been normalised to the effective radius. Targets surrounded by an open diamond are flagged as AGN by one of the indicators described in \S~\ref{subsec:agn}. The large squares correspond to the weighted average gradient in three bins of stellar mass for each redshift, the black diamonds show the weighted average for the full sample. The black line is a simple 1D fit to those points. From a sequence of 1000 Spearman correlation tests, randomly varying the sample around the uncertainties, this trend is significant at 2.8$\sigma$ in both panels. \label{fig:grad}}
\end{figure}

\npar
The final gradient measurements are reported in Table~\ref{tab:properties} in the Appendix. Most of the observed gradients are flat, only 15/180 targets exhibit a gradient significantly offset from zero (at $>3\sigma$), of which 13 gradients are negative and 2 positive. Figure~\ref{fig:grad} shows the full sample of measured gradients as a function of stellar mass in [dex/kpc] and [dex/$r_e$] in the top and bottom panels respectively, color-coded by redshift. Based on a sequence of 1000 Spearman correlation tests, where we randomly vary the sample of gradients according to their uncertainties, the negative correlation with stellar mass is significant at 2.8$\sigma$. We find a positive correlation with specific star formation rate (sSFR) at 2.5$\sigma$, the significance reduces to 1.5$\sigma$ when the sSFR is normalised to the main-sequence at each redshift. \cite{Stott2014} have similarly reported a weak positive trend with specific star formation rate at 2.9$\sigma$. No significant correlations are present with integrated [N~II]/H$\alpha$ or any of the dynamical (dispersion, velocity gradient, $v_{\mathrm{obs}}/\sigma_0$) or structural (effective radius, inclination) properties of the galaxies. 

The somewhat steeper gradients in more massive galaxies could be due to residual influence of AGN- or shock-ionisation on the [N~II] emission, since these processes are more prevalent at the high mass end. It could additionally signify stronger inside-out growth in massive galaxies \citep[e.g.][]{Nelson2015}. Finally, feedback can more easily redistribute metals in the shallower potential wells of lower mass galaxies, flattening the gradients. \cite{Stott2014} link the correlation with sSFR to the inflow of metal-poor gas flattening the gradients of strongly star-forming systems.

\begin{figure*}
\centering
\includegraphics[width=\textwidth]{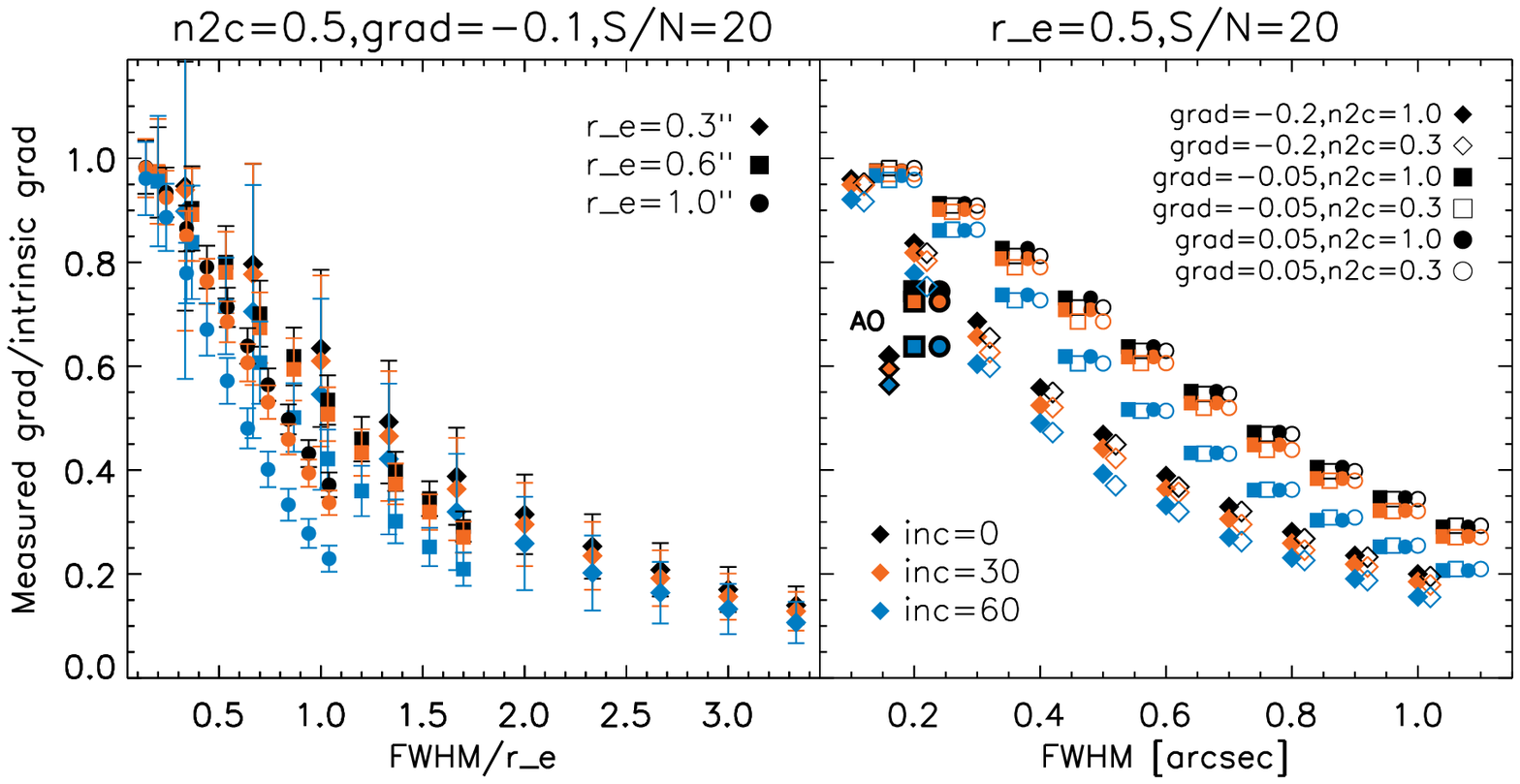}
\caption{Ratio of the recovered vs intrinsic radial $\Delta$N2/$\Delta$r gradient. The color-coding in both panels shows a range of inclinations from face-on (inc=0\degree) to almost edge-on (inc=60\degree). In the \textit{left} panel, the x-axis represents the FWHM of the observations divided by the effective radius for $r_e = 0.3, 0.6$ and 1.0\arcsec. The other parameters (central [N~II]/H$\alpha$ value (n2c), intrinsic gradient (grad) and S/N) are kept at their default values as specified in Table~\ref{tab:dysmal}. The ratios have all been derived at the same discretised set of input FWHM, but to improve the clarity of the figure we have shifted the data points for different effective radii slightly along the x-axis. Gradients are more affected by beam-smearing for larger ratios of seeing versus galaxy size. The \textit{right} panel compares the ratio of the recovered vs intrinsic gradient as a function of the FWHM of the observations for different intrinsic gradients and central [N~II]/H$\alpha$ ratios. In this panel the effective radius is kept fixed at its default value of 0.5\arcsec. The positions along the x-axis have similarly been shifted to improve the clarity of the figure. A shallower intrinsic gradient is less affected by beam-smearing, while the central N2 ratio has no impact. The black-edged symbols shown at FWHM$\sim0.16$\arcsec\ correspond to the double component PSF of SINFONI+AO data. \label{fig:beam}}
\end{figure*}

\subsection{Beam-smearing}
Before physically interpreting the observed gradients, it is crucially important to understand the effects of the coarse spatial resolution of seeing-limited data on the intrinsic abundance gradients of high-redshift galaxies. For this purpose we have analysed a set of exponential disk models created with the IDL code \textsc{dysmal} \citep{Cresci2009, Davies2011} for a range of inclination and effective radius. We add [N~II] emission assuming a smooth linear radial gradient with a chosen central [N~II]/H$\alpha$ ratio and slope. The model cubes are convolved with a Gaussian PSF for a range of assumed FWHM from 0.1\arcsec\ for some AO-assisted data to 1.0\arcsec\ for seeing-limited data taken in bad observing conditions, and random noise is added to achieve a pre-defined signal-to-noise ratio for the central pixel. Table~\ref{tab:dysmal} summarises the default value (in bold) and explored range for each parameter, chosen to reflect the range of properties found in the KMOS$^{\mathrm{3D}}$ sample. All model cubes are analysed in the exact same way as described in \S\ref{sec:grad} for the KMOS data.

\begin{table}
\centering
\caption{Parameters used for model data cubes to explore beam-smearing effects. \label{tab:dysmal}}
\medskip
\begin{tabularx}{9cm}{lc}
\toprule
FWHM~[\arcsec]   &   0.1, 0.2, 0.3, 0.4, 0.5, \textbf{0.6}, 0.7, 0.8, 0.9, 1.0 + AO \\
$r_e$~[\arcsec]     &   0.3, 0.4, \textbf{0.5}, 0.6, 0.7, 0.8, 0.9, 1.0 \\
inclination~[\degree]          &  0, 30, \textbf{45}, 60, 75 \\
S/N$_\mathrm{central}$    & 10, \textbf{20}, 30 \\
$\mathrm{[N~II]}$/H$\alpha_\mathrm{central}$    & 0.3, \textbf{0.5}, 1.0 \\
gradient~[dex/kpc] & 0.2, 0.1, 0.05, -0.05, \textbf{-0.1}, -0.2 \\ 
\bottomrule
\end{tabularx}
\end{table} 

From this first order analysis, we find that the beam-smearing mainly depends on the ratio of the FWHM of the data to the effective radius of the galaxy, which determines how well it can be resolved radially. The right panel of Figure~\ref{fig:beam} additionally shows that more highly-inclined galaxies and intrinsically steeper gradients are each affected more, as would be expected. The central [N~II]/H$\alpha$ ratio does not matter, as long as one has enough sensitivity to detect [N~II] in the outer disk. And finally, the S/N of the central pixel only impacts the uncertainties, as was also found by \cite{Yuan2013}. 

The KMOS$^{\mathrm{3D}}$ data are taken under mostly uniform seeing conditions (FWHM=$0.55\arcsec\pm0.09$\arcsec). From our models, we see that we can recover at most 70\% of the intrinsic gradient for the largest face-on disks, down to only $\sim25$\% in the smallest, highly-inclined targets. To compare beam-smearing effects with AO-assisted datasets, we also ran simulations assuming the PSF properties of SINFONI+AO observations, for which the PSF is best characterised with a double-component Gaussian model to capture the AO-corrected narrow core component with FWHM=0.16\arcsec\ and 86\% of the peak flux, and the uncorrected broad halo underneath with FWHM=0.5\arcsec\ and 14\% of the peak flux \citep[][F\"{o}rster Schreiber et~al. 2016, in prep.]{Forster2009}. The halo causes a larger beam-smearing effect than measured for a single Gaussian FWHM of 0.16\arcsec. On average, AO data recovers 65\% of the gradient, as opposed to 50\% at seeing-limited resolution. 
Utilizing the HST grism capabilities as in \cite{Jones2015}, delivers a single narrow PSF at FWHM$\sim0.2$\arcsec\ and allows the most accurate gradient measurement.

Two previous efforts to take into account the effects of beam-smearing have been made. \cite{Yuan2013} simulate high-z datacubes with different seeing and S/N values based on the measured central abundance and abundance gradient for one specific local galaxy (IRAS F17222-5953) and recover only 13\% of the intrinsic gradient when using three uniform radial bins at FWHM=0.6\arcsec. For the KMOS observations of 20 HiZELS galaxies at $z=0.8$, \cite{Stott2014} perform a simulation of 1000 disk galaxies with a range of inclinations and input gradients and recover 80\% of the intrinsic gradient in a face-on disk at 0.7\arcsec\ seeing. For an inclination of 80\degree, this reduces to 30\%. This is in contrast to our results, which only recover 35\% at FWHM=0.7\arcsec\, and are much less dependent on inclination. It remains unclear from where this difference arises. From our analysis, a large dependence on inclination is unexpected, since the use of elliptical apertures along the major axis following the axis ratio of the H$\alpha$ contours is designed to mostly take this into account.

\begin{figure*}
\centering
\includegraphics[width=\textwidth]{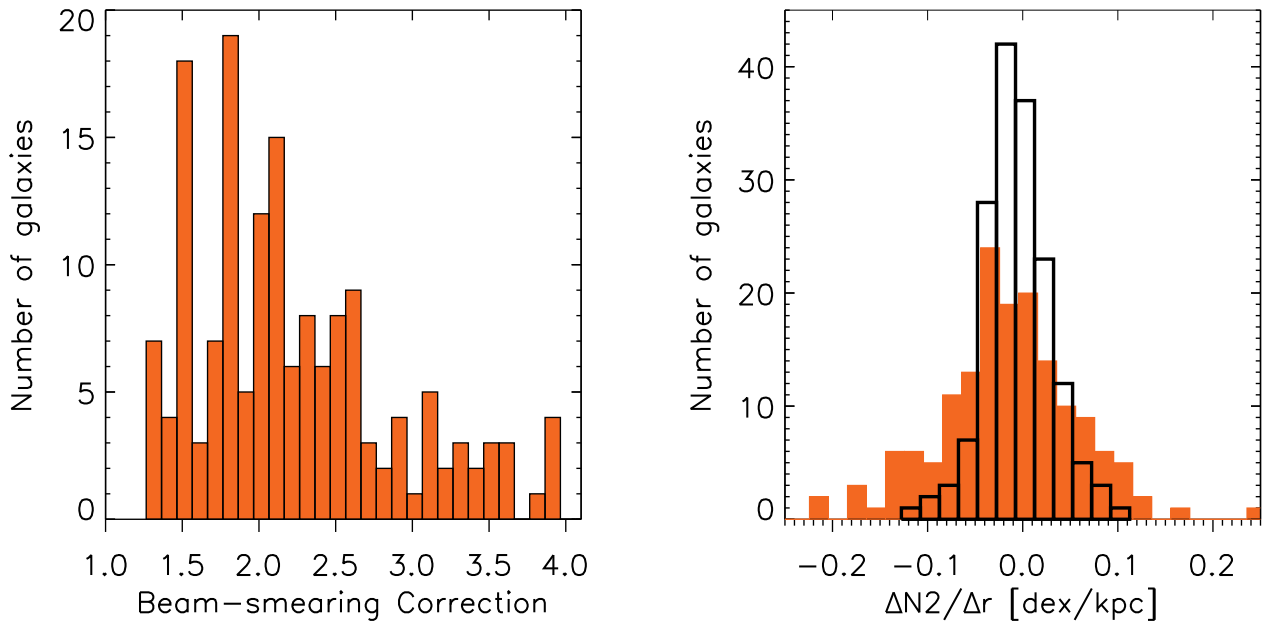}
\caption{\textit{(Left)} Histogram of the required beam-smearing corrections for the 180 gradients measured in the KMOS$^{\mathrm{3D}}$ sample. The corrections are based on the relevant seeing, and the effective radius and inclination of the targets as derived from GALFIT modelling of the F160W CANDELS photometry. \textit{(Right)} Histograms of observed (black) and intrinsic (filled orange) $\Delta$N2/$\Delta$r gradients. \label{fig:beamcorr}}
\end{figure*}

Under the assumption that we know the effective radius and inclination of each KMOS$^{\mathrm{3D}}$ target from GALFIT modelling of the F160W CANDELS photometry \citep{vanderwel2012}, and that the line emission mostly follows the stellar light as traced by the F160W continuum, we can derive the necessary beam-smearing correction from our set of models. Figure~\ref{fig:howtogetcorr} in the Appendix visualises how this is done for three examples. For a known combination of FWHM, inclination and effective radius, a certain intrinsic gradient will be beam-smeared into a unique observed gradient. We derive this match for a range of intrinsic gradients between -0.2 and 0.2~dex/kpc for each target, and then simply invert this correlation to derive the intrinsic gradient that corresponds to the actual observed gradient of the target. A histogram of the resulting beam-smearing correction factors and corrected gradients for the full sample is shown in Figure~\ref{fig:beamcorr}. We caution that each corrected gradient in this histogram has large uncertainties. We re-investigate correlations between the now-intrinsic gradients and the stellar population, structural and kinematic parameters and again do not find any statistically significant trends. Finally, we note that these corrections are derived under the assumption of a smooth, linear radial gradient. Given the clumpy nature and often irregular morphology of high redshift star-forming galaxies, this might be too strong of a simplification. In future work we will explore this in more detail with hydrodynamical simulations.

\subsection{Gradients in Interacting Systems}
\label{subsec:interacting}
In the local Universe, galaxies undergoing an interaction show flattened gradients \citep[e.g.][]{Kewley2010,Rupke2010b,Sanchez2014}. This is expected from simulations, as the interaction causes inflows of metal-poor gas from the outskirts into the galaxy centre \citep{Rupke2010b,Perez2011}. At high redshift, the situation remains unclear. For their samples of lensed galaxies, \cite{Jones2013} and \cite{Leet2015} report shallower gradients for the merging or dynamically disturbed systems. Similarly, four out of the seven galaxies with a positive abundance gradient in the seeing-limited MASSIV sample at $z\sim1.2$ are interacting \citep{Queyrel2012}. In contrast, \cite{Cresci2010} and \cite{Troncoso2014} find positive gradients in rotating disks, which would require the inflow of metal-poor gas into the galaxy centre through minor mergers, cold flows or violent disk instabilities \citep[e.g.][]{Dekel2014,Zolotov2015}.

\begin{figure}
\centering
\includegraphics[width=9cm]{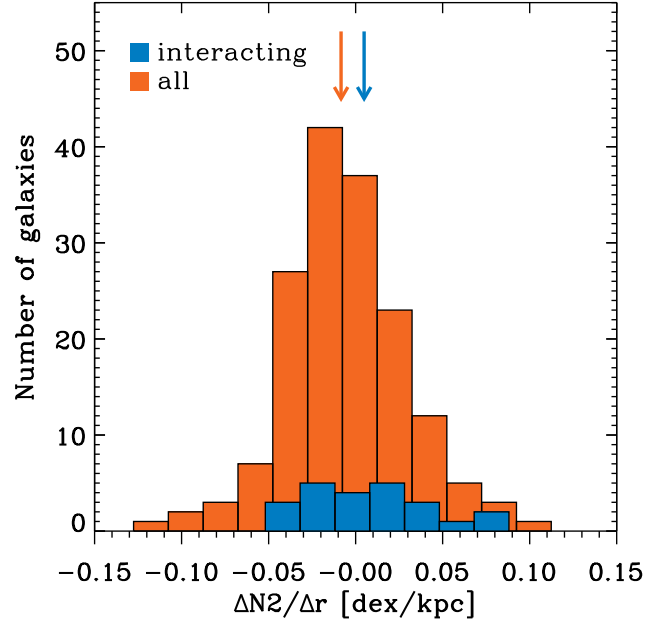}
\caption{Comparison of the full sample of KMOS$^{\mathrm{3D}}$ $\Delta$N2/$\Delta$r gradients in orange versus the 23 interacting systems in blue. The difference is not statistically significant. The orange and blue arrows indicate the median gradient of $-0.008\pm0.003$ and $0.005\pm0.008$ for the full sample and pairs respectively. Given the uncertainties, the interacting systems do not show significantly flatter gradients. \label{fig:grad_pairs}}
\end{figure}

Within the KMOS$^{\mathrm{3D}}$ sample, we find only a handful of galaxies with a statistically significant positive gradient. \cite{Wisnioski2015} define a sample of pairs based on a search for companions in the 3D-HST catalog with a velocity separation of 500~km/s, expected to fall within the IFU of the primary target - this roughly translates to a projected separation of 1.5\arcsec\ or 12~kpc. There are 39 such targets in the sample, for which we can recover 18 gradients. Additionally, one of the companion galaxies is sufficiently detected and extended for an additional gradient measurement. We also extend our definition of pairs to the more commonly used projected separation $<50$~kpc (restricted to spectroscopic redshifts), and classify 4 additional KMOS$^{\mathrm{3D}}$ targets with a gradient measurement as interacting, for a total sample of 23.
Figure~\ref{fig:grad_pairs} compares the 23 gradients in interacting systems to the full KMOS$^{\mathrm{3D}}$ sample. The median gradient of $0.005\pm0.008$ (standard deviation 0.036) is not significantly flattened compared to the full median of $-0.008\pm0.003$ (standard deviation 0.035). A Kolmogorov-Smirnov test indicates a likelihood of 29\% that both sets originate from the same parent sample. When first applying the beam-smearing correction, we find $0.011\pm0.016$ (standard deviation 0.08) for the pairs and $-0.015\pm0.008$ (standard deviation 0.10) for the full sample, with a KS-test likelihood of 31\%.

\begin{figure*}
\centering
\includegraphics[width=\textwidth]{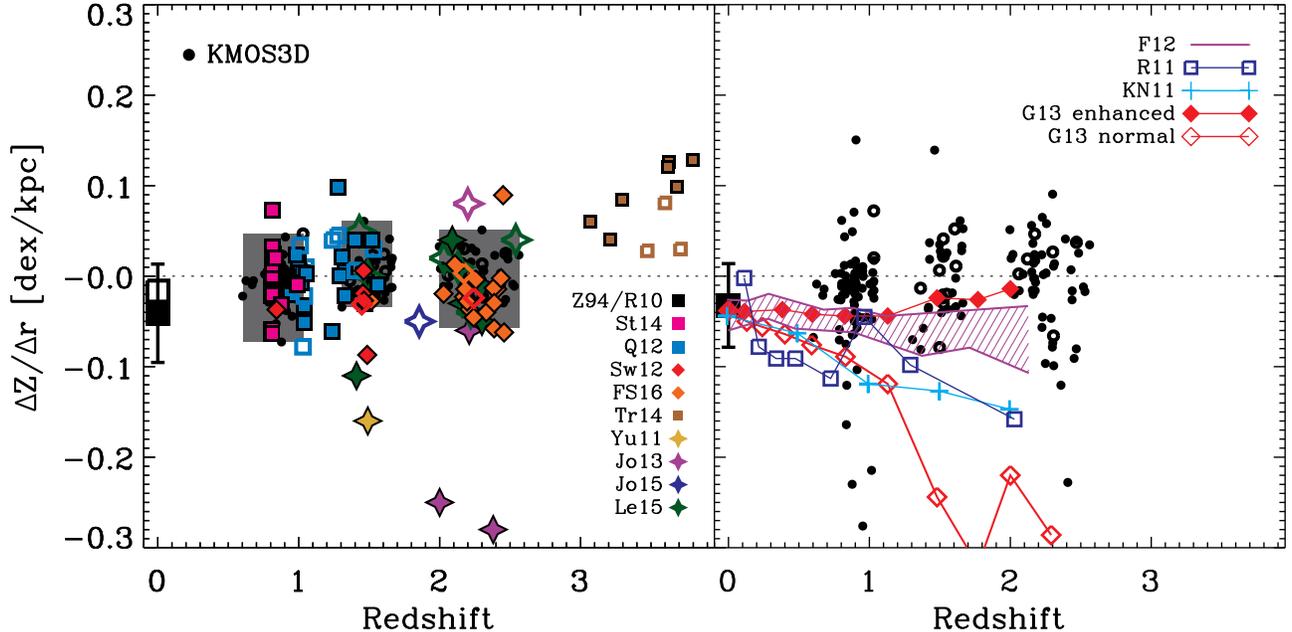}
\caption{The left panel compiles the KMOS$^{\mathrm{3D}}$ results with all the abundance gradients measured at high redshift from the literature. Open symbols refer to interacting or disturbed systems, as defined in the various papers. In the local Universe, the filled/open black square shows the median gradient for a sample of isolated spiral galaxies \citep{Rupke2010b,Zaritsky1994} and interacting spiral systems \citep{Rupke2010b} respectively. The star symbols refer to gravitationally lensed galaxies \citep{Yuan2011, Jones2013, Jones2015, Leet2015}; diamonds have AO-assisted data \citep[][F\"{o}rster Schreiber et~al. 2016 (in prep.)]{Swinbank2012}; squares have only seeing-limited data \citep{Queyrel2012,Stott2014,Troncoso2014}. KMOS$^{\mathrm{3D}}$ data is represented by small black open or closed circles. To aid visibility, the grey background squares indicate the spread of the KMOS$^{\mathrm{3D}}$ measurements in the three redshift intervals. 
In the \textit{right} panel, the KMOS$^{\mathrm{3D}}$ galaxies have been corrected for the effect of beam-smearing. We additionally plot different theoretical predictions for the time evolution of abundance gradients from \cite{Pilkington2012} based on simulations in \cite{Rahimi2011,Kobayashi2011,Few2012,Gibson2013}. \label{fig:grad_zz}}
\end{figure*}

\subsection{Evolution of Metallicity Gradients with Cosmic Time}
\label{subsec:grad_evolution}
Figure~\ref{fig:grad_zz} compiles available abundance gradients from the literature as a function of redshift. For this compilation, we convert our gradients in [N~II]/H$\alpha$ ($\Delta$N2/$\Delta$r) into abundance gradients $\Delta$Z/$\Delta$r by multiplying with a factor 0.57 in accordance with the linear conversion by \cite{pp04}. In the local Universe, the median gradient for a sample of isolated spiral galaxies from \cite{Rupke2010b} and \cite{Zaritsky1994} is -0.04$\pm$0.05~dex/kpc. The open square symbol shows the flatter gradient of $-0.0165$~dex/kpc for galaxies currently undergoing an interaction \citep{Rupke2010b}. These studies use the $R_{23}$ metallicity indicator of ([O~II]~$\lambda\lambda$~3727,3729 + [O~III]~$\lambda$4959 + [O~III]~$\lambda$5007)/H$\beta$. \cite{Rupke2010b} compare their results to abundances calculated from [N~II]~$\lambda$6583/[O~II]~$\lambda\lambda$~3727,3729 and ([O~III]~$\lambda$5007)/H$\beta$)/([N~II]~$\lambda$6583/H$\alpha$) (the O3N2 indicator), and find no systematic difference after calibrating the different indicators following \cite{Kewley2008}. At high redshift, abundance gradients are almost exclusively derived from the [N~II]/H$\alpha$ ratio, since it only requires observations in one band and is insensitive to dust correction. \cite{Jones2013} and \cite{Leet2015} additionally measure O3N2 for a subset of their samples and find consistent results. The grism data from \cite{Jones2015}, and the $z>3$ sample from \cite{Troncoso2014} do not have access to H$\alpha$, and use the R23 diagnostic instead. For the literature sample in Figure~\ref{fig:grad_zz}, the star symbols correspond to lensed galaxies with AO, diamonds are non-lensed galaxies with AO, and squares show seeing-limited data. Open symbols show interacting or disturbed systems, as defined in the various papers from kinematic classification, visual inspection of broad-band imaging, or a combination of both. For KMOS$^{\mathrm{3D}}$, we use the pair classification as described above.

With the availability of larger samples, the steep abundance gradients found in lensed galaxies by \cite{Jones2013} and \cite{Yuan2011} that spurred much of the initial debate on the feasibility of gradient measurements in field galaxies, as well as the idea of a strong flattening of gradients over cosmic time, now seem more like outliers. These lensed galaxies have lower stellar masses than typically targeted in the other surveys (the three steepest gradients in Figure~\ref{fig:grad_zz} were measured for $\log(M_*/\mathrm{M}_\odot)=9.7$ and 9.9 \citep{Jones2013} and a dynamical mass of 10.42~M$_\odot$ \citep{Yuan2011}), but there is still overlap. The KMOS$^{\mathrm{3D}}$ survey for example probes down to $\log(M_*/\mathrm{M}_\odot)=9.7$, as can be seen in Figure~\ref{fig:grad}. Also, the larger study of lensed galaxies in \cite{Leet2015} did not uncover any additional steep gradients. 
\npar
In the right panel of Figure~\ref{fig:grad_zz}, the KMOS$^{\mathrm{3D}}$ gradients are corrected for beam-smearing. We can not extend this correction to the literature data, since we do not always know the FWHM, effective radius and inclination of the targets. Overplotted are models for the evolution of gradients over time from \cite{Rahimi2011,Kobayashi2011,Few2012} and \cite{Gibson2013} as derived by \cite{Pilkington2012}. The red open and filled diamonds compare the normal and enhanced feedback mode for the same simulation from \cite{Gibson2013}. There is large scatter in the observed gradients, and remaining uncertainty with regards to the beam-smearing correction, but based on these models, the scarcity of gradients much steeper than -0.1~dex/kpc at $z\sim2$ suggests the need for strong feedback.

\section{Summary and Discussion}
\label{sec:summ}
This work takes advantage of the large sample size and wide redshift coverage of the KMOS$^{\mathrm{3D}}$ near-IR IFU survey to infer the integrated metal abundance and abundance gradient of star-forming galaxies at the peak of cosmic star formation activity between $z=0.6-2.6$ from the [N~II]/H$\alpha$ flux ratio. Such measurements can shed crucial light on the main physical processes driving the assembly of stellar mass and heavy elements in the Universe. Our main conclusions are as follows:
\begin{itemize}
\item Our mass-metallicity relation confirms previous literature results. Excluding AGN reduces sensitivity at the high-mass end, but does not significantly change the relation at lower masses ($\log(M_*/\mathrm{M}_\odot)\lesssim10.8$). We do not find a significant effect of SFR, sSFR or galaxy environment on the observed [N~II]/H$\alpha$ ratios.
\item A thorough analysis of the effect of beam-smearing in seeing-limited data at high redshift shows that one can recover between 30\% and 70\% of the intrinsic abundance gradient, depending mostly on the ratio of the FWHM of the data to the effective radius, and to a lesser extent on galaxy inclination.
\item The majority of observed [N~II]/H$\alpha$ gradients are flat. Cosmological simulations suggest the need for strong feedback to create flat abundance gradients at high redshift. No statistically significant correlations with stellar, kinematic or structural galaxy properties are found.
\end{itemize}

A number of factors could cause intrinsically flat abundance gradients. One effect is the radial inflow of metal-poor gas and metal-mixing caused by a merger. In the local Universe, this has been established to cause shallower gradients, both observationally and in simulations \citep{Rupke2010a, Rupke2010b}. At high redshift, the merger fraction remains highly debated with most observational results pointing to consistent or higher fractions than found locally \citep[e.g.][]{Lotz2008,Man2012,Lopez2014,Lackner2014}. However, not all high-z surveys systematically find shallower gradients for disturbed systems, as discussed in Sections~\ref{subsec:interacting} and \ref{subsec:grad_evolution}. 

A second important physical process is feedback, which can blow out metal-rich gas from the centre and redistribute it at larger radii. The ubiquitous presence of both AGN and star formation driven feedback is well established at high redshift, especially at high mass and star formation surface density \citep[e.g.][]{Newman2012,Genzel2014}. 
\npar
Aside from these physical effects, one has to take into account the various reasons why the \textit{observed} gradient could be flat. Beam-smearing is the most important factor here, and while we have attempted to address this, significant uncertainties remain. Secondly, in analogy with local SFGs, the fraction of warm diffuse ionised gas increases in radial apertures further out in the galaxy disks, and could strongly bias the measured line ratios there \citep{Yuan2011}. Adaptive-optics assisted data reaches the resolution of massive star-forming clumps at $z\sim2$ and could thus minimise this effect, but the uncorrected seeing-limited halo still contributes to the beam-smearing correction (Figure~\ref{fig:beam}). Space-based grism data is more suited, but struggles with sensitivity. 

Similar biases are introduced when shock fronts alter the photo-ionization conditions of the interstellar medium \citep[e.g.][]{Kewley2013,Newman2014}. These biases affect any abundance indicator based on rest-frame optical emission line ratios, but have especially significant effects for the N2 indicator. Spatially resolved coverage of multiple lines would allow better rejection of spatial pixels influenced by AGN or shock ionisation based on the BPT diagram. 


\npar
This analysis has highlighted the need for high-quality data with sufficient spatial resolution, S/N and  wavelength coverage to disentangle the physical processes driving the observed line ratios and to pin down the intrinsic abundance gradients. Our understanding of the global and resolved metallicity distributions among and within SFGs, as well as their cosmic evolution, will benefit greatly from the added resolution and sensitivity of JWST and the ELTs. 


\appendix
\begin{figure*}
\centering
\includegraphics[width=\textwidth]{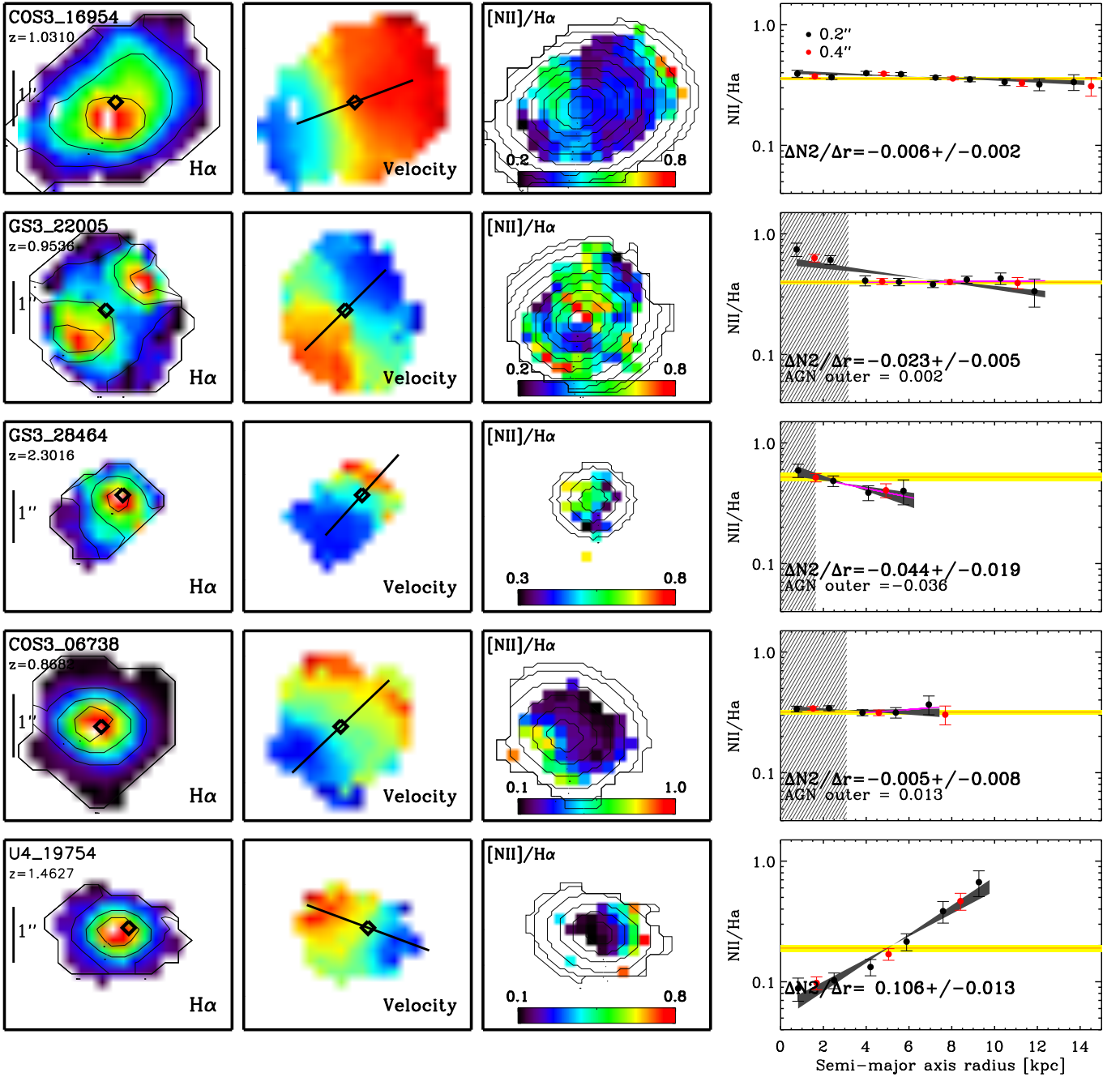}
\caption{Illustration of how abundance gradients are measured for five example galaxies. The three panels on the left show the H$\alpha$, velocity and [N~II]/H$\alpha$ maps respectively, where in the latter only pixels with S/N$>3$ in [N~II] are shown. In each panel, the black open diamond corresponds to the continuum centre. The H$\alpha$ map additionally shows the H$\alpha$ contours, and the kinematic position angle is shown on the velocity map. Based on the centre, PA and ellipticity of the outer H$\alpha$ contours, we create one-pixel wide elliptical apertures as shown on top of the [N~II]/H$\alpha$ map in the third column. The chosen ellipticity does not significantly affect the measured gradient. The rightmost panels show the [N~II]/H$\alpha$ ratios as a function of semi-major axis radius for these apertures. The red data points correspond to the wider 2 pixel or 0.4\arcsec\ apertures. The integrated [N~II]/H$\alpha$ ratio and its uncertainty are indicated by the orange horizontal line and yellow band. 
The dark grey shaded band is the best-fit gradient and uncertainty. The middle 3 galaxies are each identified as an AGN. In this case we exclude the inner one or two apertures as indicated by the vertical grey shaded region, and refit the remaining outer data points (magenta line). For the third and fourth galaxy, both fits are consistent. The second row shows an example where the [N~II]/H$\alpha$ ratios in the centre are significantly elevated due to the AGN, such that the outer gradient is much flatter.\label{fig:n2grad_examples}}
\end{figure*}

\begin{figure*}
\centering
\includegraphics[width=\textwidth]{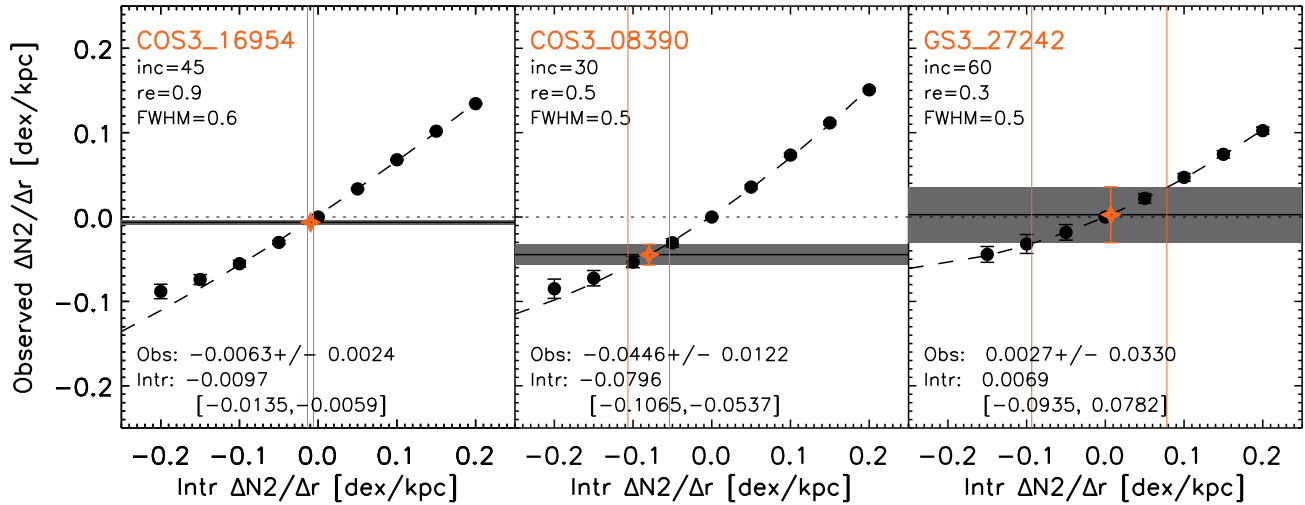}
\caption{This figure provides three examples of the beam-smearing correction and intrinsic metallicity gradient. In each panel, the horizontal grey band and black line show the observed gradient and its uncertainty. The black circles correspond to the gradient we would have observed for a range of intrinsic gradients from -0.2 to 0.2~dex/kpc, given the known seeing, inclination and effective radius of the target. We fit a polynomial to this correlation, and invert it to derive the intrinsic gradient and uncertainties which corresponds to the gradient we actually observed in the data. This is shown by the orange star and orange vertical lines. The observed and intrinsic gradients are noted in [dex/kpc] in the bottom of each panel. In the right panel, the relation between observed and intrinsic gradient is very flat due to the small effective radius and high inclination of the target, requiring a large correction. This additionally causes large uncertainties on the intrinsic gradient compared to the observed uncertainties. \label{fig:howtogetcorr}}
\end{figure*}

\begin{acknowledgments}
MF and DW ackowledge the support of the Deutsche Forschungs Gemeinschaft (DFG) via Project WI 3871/1-1.
\end{acknowledgments}


\begin{table}
\centering
\caption{Target properties. \label{tab:properties}}
\medskip
\begin{tabularx}{11cm}{rrcrrcrl}
\multicolumn{1}{l}{\#} & \multicolumn{1}{c}{$z$} & \multicolumn{1}{c}{$\log(M_*/\mathrm{M}_\odot)$} & \multicolumn{1}{c}{SFR$_{\mathrm{H}\alpha}$\tablenotemark{a}} & \multicolumn{1}{c}{[N~II]/H$\alpha$\tablenotemark{b}} & \multicolumn{1}{c}{$\Delta N2 / \Delta r$} & \multicolumn{1}{r}{beam-smearing}\\ \toprule
\multicolumn{1}{l}{} & \multicolumn{1}{c}{} & \multicolumn{1}{c}{} & \multicolumn{1}{c}{[M$_\odot$/yr]} & \multicolumn{1}{c}{} & \multicolumn{1}{c}{[dex/kpc]} & \multicolumn{1}{r}{}\\ \midrule

           1   &   0.758   &   10.66   &      5.5   &   $0.41\pm0.03$   &   $ 0.004\pm0.016$   &   1.3\\
           2   &   0.775   &   10.94   &      0.9   &   $<0.11$   &                     &      \\
           3   &   0.777   &   10.35   &     24.7   &   $0.44\pm0.03$   &   $-0.013\pm0.015$   &   3.4\\
           4   &   0.787   &   10.78   &      4.6   &   $0.56\pm0.03$   &   $-0.038\pm0.013$   &   3.3\\
           5   &   0.787   &   10.51   &     21.0   &   $0.41\pm0.03$   &   $-0.050\pm0.014$   &   2.1\\ \bottomrule
\tablenotetext{NOTE}{This Table is published in its entirety in the electronic edition of the Astrophysical Journal. A portion is shown here for guidance regarding its form and content.}
\tablenotetext{a}{H$\alpha$-based SFR corrected for dust extinction using the SED-derived reddening and accounting for additional extinction of the nebular lines following Wuyts et al. (2013)}
\tablenotetext{b}{Integrated [N~II]/H$\alpha$ ratio}
\tablenotetext{c}{Observed gradient in [N~II]/H$\alpha$, excluding the nuclear region for AGN-flagged targets. Reported only when measurable.}
\tablenotetext{d}{Multiplicative correction factor for the N2 gradient.}
\end{tabularx}
\end{table}

\end{document}